\documentclass[11pt,aps,prd,notitlepage,longbibliography]{revtex4-2}
\usepackage{rotating}% Rotate figures
\usepackage{amsmath}
\usepackage{amssymb}
\usepackage{mathrsfs}
\usepackage{bm}
\usepackage{xcolor}

\usepackage{graphicx}
\usepackage{hyperref}
\hypersetup{
    colorlinks = true,
    urlcolor   = blue,
    citecolor  = black,
}

\newcommand{\RomanNumeralCaps}[1]
\linenumbers

\makeatother

\newcommand{\beq}{\begin{equation}}
\newcommand{\eeq}{\end{equation}}

\def\A{\boldsymbol{A}}

\def\D{\boldsymbol{D}}
\def\E{\boldsymbol{E}}
\def\f{{\boldsymbol f}}

\def\I{{\bf I}}

\def\p{\boldsymbol{p}}

\def\u{\boldsymbol{u}}
\def\k{\boldsymbol{k}}

\def\x{\boldsymbol{x}}

\def\I{{\bf I}}

\def\0{{\bf 0}}

\def\bsig{{\bm \sigma}}
\def\bSig{\bm \Sigma}

\emergencystretch=\maxdimen
\begin{document}

\title{Hydrodynamic instabilities and collective dynamics in activity-balanced pusher-puller mixtures}

\author{Bryce Palmer$^{1}$}
\author{Wen Yan$^{3}$}
\author{Tong Gao$^{1,2}$}
\email{gaotong@egr.msu.edu}
\affiliation{
$^1$Department of Mechanical Engineering, Michigan State University, East Lansing, MI 48864\\
$^2$Department of Computational Mathematics, Science and Engineering, Michigan State University, East Lansing, MI 48864 \\
$^3$Center for Computational Biology, Flatiron Institute, Simons Foundation, New York, New York 10010
}

\date{\today}

\begin{abstract}
Microorganisms living in microfluidic environments often form multispecial swarms, where they can leverage collective motions to achieve enhanced transport and spreading. Nevertheless, there is a general lack of physical understandings of the origins of the multiscale unstable dynamics observed within these systems. Here, we build a computational model to study binary suspensions of rear- and front-actuated microswimmers, or respectively the so-called ``pusher'' and ``puller'' particles, that have different populations and swimming speeds. We perform direct particle simulations to reveal that collective system dynamics are possible even in the scenario of an ``activity-balanced'' mixture, which produces near zero mean extra stress. We first construct a continuum kinetic model to describe the initial transient period when the system is near uniform isotropy and then perform linear stability analysis to reveal the system's finite-wavelength hydrodynamic instabilities, in contrast with the long-wavelength instabilities of pure pusher/puller suspensions. Then, we carry out slender-body discrete particle simulations to resolve both the short time instabilities and the the longtime dynamics, which feature non-trivial density fluctuations and spatially-correlated motions, distinct from those of single-species.
\end{abstract}
\maketitle

\section{Introduction}
Active suspensions of swimming microorganisms, such as bacteria or algae, can exhibit fascinating collective behaviors that feature large-scale coherent structures, enhanced mixing, ordering transition, and anomalous diffusion \cite{ramaswamy10,Shelley2016}. In the limit of vanishing Reynolds numbers, self-driven or swimming micro-particles effectively exert stresses upon the ambient liquid, which act as a coupling medium for the generation of large-scale collective dynamics.
To uncover the multiscale origins of these unstable dynamics, researchers have constructed micro-mechanical models that can capture the distinctive swimming mechanisms for various microswimmers. Figure 1(a) sketches the so-called ``pusher'' and ``puller'' particles to characterize a class of microswimmers that respectively generate thrust at the rear (e.g., {\it E. Coli}) and at the front (e.g., {\it Chlamydomonas}) of the body \cite{Ishikawa06}. The swimming motions of such rodlike particles can produce dipolar extra stresses, the strength of which can be either negative (i.e., extensile) in the case of pushers or positive (i.e., contractile) in the case of pullers. The magnitude of these extra stresses scales as $\mu u_c \ell^2$ where $\mu$ is a fluid viscosity, $u_c$ is a certain characteristic swimming speed, and $\ell$ is the rod length \cite{saintillan08a}.

Thus far, our physical understanding of collective dynamics observed in active suspensions has been largely limited to the idealized situation of a single particle species. In the dilute limit, it is well understood that pure pusher suspensions exert non-zero mean (or net) extensile stresses, which drive large-scale active flows; on the other hand, net contractile stresses produced by pure puller suspensions suppress the development of active flows  \cite{ramaswamy10,saintillan08a}.
Nevertheless, in real-world ecosystems, multiple species of microorganisms may coexist with competition and coordination due to biological, chemical, and social mechanisms; these mixtures often utilize collective motions to achieve efficient locomotion or spreading \cite{BenJacob16}. Several ``dry'' models that neglect hydrodynamic interactions have been used to study non-equilibrium phase transition and pattern formation of multispecies swarms \cite{wittmann_effective_2018, maloney_clustering_2020, stenhammar_activity-induced_2015, kolb_active_2020, takatori_theory_2015}. Yet, far less is known about the role of fluid mechanics when different types of micro-swimmers interact collectively. Among the minimal studies of ``wet'' systems \cite{pessot18,dora2020}, we learn that the collective dynamics of pure pushers can be suppressed by adding pullers, or equivalently, adding contractile stresses, which effectively reduce or ``neutralize'' the extensile stresses produced by pushers. However, these studies only consider the simplest scenario of dilute binary suspensions where pushers and pullers have the same shape, population size, and swimming speed. In this scenario, the mean stress of the entire mixture exerted on the ambient fluid depends solely on the relative concentrations of the two species, as members of opposing species produce equal and opposite extra stresses.

In this study, we construct a hybrid computational model that integrates a continuum model and discrete particle dynamics to investigate the unstable dynamics of a dilute pusher-puller mixture where the two types of microswimmers can have different swimming speeds and particle concentrations. However, we define the system to be ``activity balanced'' such that the isolated stresslet produced by the two species exactly cancel one another, meaning that the ratio of each species' isolated swimming speed is inversely proportional to the ratio of their number fraction. Compared to previous single- or multi-species models, where a net extensile extra stress can be used as an indicator of instabilities and collective motions \cite{saintillan08a,Brotto15}, our system provides a unique framework whereby one can investigate the underlying physics of binary suspensions in the absence of global, activity-driven extra stresses. For dilute suspensions, this constraint effectively neutralizes the leading order system stress leaving only higher order stress fluctuations due to many-body interactions. Our key findings suggest that starting with zero mean extra stress, the interplay of the competing mechanisms, induced by differences in speed and relative particle number fractions, can produce intriguing collective dynamics at late times while the system remains near zero mean extra stress. Unlike the classical ``activity-unbalanced'' cases that feature strongly correlated active flows with large density and velocity fluctuations, here the activity-balanced cases typically exhibit weaker and moderately correlated motions with non-trivial velocity and concentration fluctuations. In contrast with other activity-balanced pusher-puller mixtures models with a $1:1$ ratio where all particles swim at the same speed \cite{dora2020}, we investigate the entire phase space of particle speeds and number densities. In agreement with these models, we find that the $1:1$ ratio case is stable and additionally acts as a critical point in the transition between two distinct types of finite-wavelength instabilities.

The paper is organized as follows. In Section \ref{sec:kinetic}, we introduce the micromechanical models of two types of microswimmers and set up the activity-balanced condition for binary suspensions. Then, we construct a mean-field kinetic model that describes the short-time behavior of such dilute binary suspensions and perform linear stability analyses. Section \ref{simu} is dedicated to the discrete particle simulations built upon the slender-body theory, which we utilize to unveil early- and late-time dynamics. To achieve this, we perform linear stability analysis on the initial transient and systematically explore how particle statistics are influenced by the model parameter space. Finally, we summarize and draw conclusions in Section \ref{sec:conclusion}.

\section{Mean-field kinetic model} \label{sec:kinetic}
\subsection{Micromechanical Model} \label{sec:micromechanical}
\begin{figure}[t]
  \centering
  \includegraphics[width=0.8\textwidth]{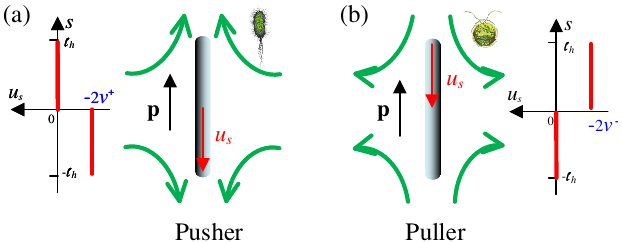}
  \caption{Schematics for a single pusher (e.g., {\it E. Coli}) and puller (e.g., {\it C. reinhardtii}) particle using the local slender body model. The green arrows represent the resultant near-body flow directions when imposing the slip velocity $u_s$ (red arrow) along the rod's center-line. In this work we model $u_s$ to be a simple step function.}
\label{fig_schematic}
\end{figure}

Consider a dilute suspension of $N$ rigid, slender rodlike active particles in a cubic periodic volume with width $L$, each of equal length $\ell$ and diameter $b$ with inverse aspect ratio $A=b/l<<1$. Among these particle, there are $N^{+}$ pushers with isolated swimming speed $v^{+}$ and $N^{-}=N-N^{+}$ pullers with speed $v^{-}$. For simplicity, we let $\ell_h=\ell/2$ be the particle half-length and denote pushers with superscript ${}^+$ and pullers with superscript ${}^-$. Using the slender body theory for the Stokes equation \cite{batchelor70,keller76}, we model their different swimming behaviors by respectively imposing a centerline, backward slip velocity according to the simple Heaviside step function $u^{\pm}_s(s) = - v^{\pm} \left[1 \mp \rm{sgn}(s)\right],~v^{\pm}>0$, as seen in Fig.~\ref{fig_schematic}. Following the classical slender body theory, we denote the dipolar stress created by a single, isolated particle with orientation $\p$ ($|\p| = 1$) as $\sigma_0^{\pm} \p\p$ where the coefficient, or the so-called ``stresslet'', $\sigma_0^{\pm} \propto \mu v^{\pm} \ell^2$ has a unit of force times length \cite{saintillan08a}. According to previous studies of active suspensions, both theory and experiments have proved that a dilute, pure pusher system with $\sigma_0^{+}<0$ can exhibit activity-driven hydrodynamic instabilities \cite{simha02,saintillan08a}. Finally, to facilitate analysis, we introduce the universal characteristic velocity scale $u_c = U_p$ ($U_p \sim 1 \mu m/s$ is a certain typical isolated rod speed) for the entire mixture and the (total) particle number density $n = N/V$, from which we obtain the following parameters: the speed ratio $\gamma^{\pm}=v^{\pm}/u_c$, the number fraction $\chi=N^+/N$ for pushers and $1-\chi=N^-/N$ for pullers.

\subsection{Governing Equations}
We start by constructing a minimal continuum model, aiming to capture the key flow physics during the initial transient period of the particle simulations where the system dynamics just start to deviate from a uniform isotropy. Performing linear stability analysis on this model can shed light upon the physical mechanisms that give rise to the onset of correlated motions in particle simulations. Here, we choose to extend the classical mean-field kinetic model for a single-species \cite{saintillan08a} to the case of a binary suspension.
To begin with, we describe the continuum configuration of the mixed species using the probability distribution functions (PDFs) $\Psi^{\pm}(\x,\p,t)$ in terms of the rods' center-of-mass (C.O.M.) position $\x$ and orientation $\p$, with the superscript ``$\pm$'' consistent with micromechanical model above. As an expression of the conservation of particle number, one can derive the Smoluchowski equations
\begin{equation}
    \frac{\partial\Psi^{\pm}}{\partial t}+\nabla\cdot(\dot{\x}^{\pm}\Psi^{\pm})+\nabla_{\p}\cdot(\dot{\p}^{\pm}\Psi^{\pm})=0
\label{smolu}
\end{equation}
where $\nabla$ is the regular spacial gradient operator and $\nabla_{\p} = (\I - \p \p)\cdot\partial/\partial \p$ is the orientational gradient operator on the unit
sphere. For both species, their PDFs satisfy the global constraints
\beq
    \frac{1}{V} \int_{V} \int_{S} \Psi^{+}(\x, \p, t) d S_p d V = \chi, \quad
    \frac{1}{V} \int_{V} \int_{S} \Psi^{-}(\x, \p, t) d S_p d V =1-\chi.
    \label{global}
\eeq
We also define the zeroth (i.e., concentration) and the second moment of $\Psi$ as
\beq
c^{\pm}(\x,t) = \int_S \Psi^{\pm}(\x,\p,t)d S_p, \quad
\boldsymbol{D}^{\pm}(\x,t) = \int_S \Psi^{\pm}(\x,\p,t) \p\p d S_p.
\eeq
Then we can derive the conformational fluxes $\dot\x^{\pm}$ and $\dot \p^{\pm}$ using local slender-body theory as
\begin{gather}
    \dot{\x}^{\pm} = \gamma^{\pm} \p+\u-d_T^{\pm} \nabla(\ln \Psi^{\pm}), \label{xdot}\\
    \dot{\p}^{\pm} = (\boldsymbol{I}-\p\p) \cdot\nabla\u \cdot \p-d_{R}^{\pm} \nabla_{\p}(\ln \Psi^{\pm}). \label{pdot}
\end{gather}
Equations \eqref{xdot}-\eqref{pdot} state that an individual particle can move at a speed of $\gamma^{\pm} \p$ relative to the background fluid velocity $\u$ and simultaneously rotate under the fluid shear by following Jeffrey's orbit \cite{jeffery22}.
These equations are nondimensionalized according to the universal characteristic velocity scale $u_c$ and length scale $\ell_c = \left(n\ell^2\right)^{-1}$ where we adopt the same notations of the dimensionless parameters $\gamma^{\pm}$ (speed ratio) and $\chi$ (relative number fraction) in Section \ref{sec:micromechanical}. In addition to the deterministic motions, diffusion effects in binary suspensions are incorporated via the dimensionless isotropic translational ($d_T^{\pm}$) and rotational ($d_R^{\pm}$) diffusion coefficient. It is worthwhile mentioning that this model also somewhat resembles the ``cyclic swimmers'' model proposed in Ref~\cite{Brotto15} in mimicking a class of particles undergoing transitions between pusher and puller states.

The above equations are closed by solving the incompressible fluid velocity $\u$ that is produced by the extra-stress tensor
$\bSig$, which is, in turn, produced by the particles' presence via a forced Stokes equation
\begin{gather}
\nabla p -\nabla^{2} \u=\nabla \cdot \bSig, \\
\nabla \cdot \u=0
\end{gather}
where $p$ is the fluid pressure. The extra stress includes both species' contributions and sums the single stresslet $\sigma^{\pm}_0 \p\p$ for all particles. Then, we follow the Kirkwood theory \cite{doi88} to derive the two configurational-averaged stresses as
\beq
\bSig^{\pm}(\x,t) = \mp \alpha \gamma^{\pm}\int_S \Psi^{\pm}\p\p d S_{p} = \mp \alpha \gamma^{\pm} \D^{\pm}(\x,t)
    \label{stress1}
\eeq
with the strength coefficients $\alpha\gamma^{\pm} = \frac{\sigma^{\pm}_0}{u_c\mu \ell^2} > 0$, leading to the total stress
\beq
\bSig =\chi \bSig^{+} + \left(1-\chi\right) \bSig^{-} = -\alpha \gamma^+ \chi \D^+(\x,t) + \alpha\gamma^- \left(1-\chi\right) \D^-(\x,t).
\label{totstress}
\eeq
In deriving Eqs.~\eqref{stress1} and \eqref{totstress}, we use the fact that the stress generated by a pusher and puller $\sigma^{\pm}$ scales with $\mu v^{\pm} \ell^{2}$, which is proportional to the single rod speed $v^{\pm} = \gamma^{\pm} u_c$. Hence, the non-dimensionalization of stress produces a different definition for $\alpha$ for a mixed species than for a single species because the non-dimensional isolated swimming speed is $\gamma^{\pm}$, rather than 1. Note that defining local total stress in \eqref{totstress} requires applying appropriate ensemble averaging in a certain small representative elementary volume, whose size needs to be selected based on the characteristic features of the bulk material properties. As well, during the initial transient period, when the system dynamics just start to deviate from the initial uniform isotropic state, $\bSig$ in \eqref{totstress} can be defined locally. We also assume that the binary suspension is dilute, i.e. we consider contributions due to short-range steric interactions (e.g., collisions) to be negligible.
%,since it may drive the isotropic-nematic phase transitions to form strong alignment structures at high particle concentrations \cite{doi88,ESS2013}.

To derive the activity-balance condition from the above governing equations, we first perturb the initial equilibrium solution of an isotropic state by incorporating the disturbance solutions that fluctuate around their global means, i.e., $\Psi^{\pm} = 1/4\pi + \epsilon\Psi'_{\pm}$ and $\u = \epsilon \u'$ ($|\epsilon|\ll1$). Here, we use superscript ``${}_{'}$'' to denote all high-order disturbance solutions hereinafter. We can calculate the following near-isotropy expansions
\begin{gather}
 \frac{c^+(\x,t)}{\chi} = 1+ \epsilon c'_{+}(\x,t), \quad \frac{c^-(\x,t)}{1-\chi} = 1+ \epsilon c'_{{-}}(\x,t)
    \label{diluteC} \\
    \D^{\pm} (\x,t) = \frac{\I}{3} + \epsilon \D'_{\pm}(\x,t).
    \label{diluteD}
\end{gather}
Then the stress defined in Eq.~\eqref{totstress} can be written as
\begin{equation}
   \bSig (\x,t) = -\alpha \left(\gamma^{+}\chi - \gamma^{-}\left(1-\chi\right)\right) \I + \epsilon\bSig'.
   \label{stress}
\end{equation}
The prefactor $-\alpha \left(\gamma^{+}\chi - \gamma^{-}\left(1-\chi\right)\right)$ essentially characterizes the mean activity for a homogeneous system. Thus, eliminating the extra stress at the leading order yields the activity-balanced condition
\begin{equation}
    \chi=\frac{\gamma^{-}}{\gamma^{+}+\gamma^{-}}, \quad 1-\chi=\frac{\gamma^{+}}{\gamma^{+}+\gamma^{-}}, \quad \chi \in \left(0,1\right).
    \label{stresscondition}
\end{equation}

\subsection{Linear Stability Analysis}
It is straightforward to perform the linear stability analysis for the governing equations \eqref{smolu}-\eqref{totstress} when subjected to the activity-balanced condition \eqref{stresscondition}. With the $\chi-\gamma^{\pm}$ relation in \eqref{stresscondition}, we can perturb the isotropic base-state PDF solutions with disturbance functions $\Psi'_{\pm}$ as
\begin{gather}
\Psi^{\pm} = \frac{\gamma^{\mp}}{4\pi\left(\gamma^{+}+\gamma^{-}\right)}\left(1+\epsilon \Psi'_{\pm}\right),
\end{gather}
which, together with the distance fluid velocity $\u'$, lead to the linearized Smoluchowski equations
\beq
    \frac{\partial\Psi'_{\pm}}{\partial t} = - \gamma^{\pm}\p\cdot\nabla\Psi'_{\pm}+ d_T^{\pm}\nabla^2\Psi'_{\pm}+3\p\p : \E',
\eeq
with $\E'=\left(\nabla \u'+\nabla \u'^{T}\right)/2$ the strain-rate tensor.
Then, we follow Ref. \cite{saintillan08a} to apply a plane-wave decomposition to the disturbance functions such that
\begin{gather}
\Psi^{\prime}_{\pm}(\x, \p, t)=\widetilde{\Psi}^{\pm}(\p, \k) \exp (i \k \cdot \x+\sigma t), \\
\u'(\x, t)=\widetilde{\u}(\k) \exp (i \k \cdot \x+\sigma t), \\
c'(\x, t)=\widetilde{c}(\k) \exp (i \k \cdot \x+\sigma t),
\label{plane-wave}
\end{gather}
where $\k$ is the wave vector and $\sigma$ is the growth rate. Both scalar and vector Fourier coefficients that are denoted with ``$\sim$'' can then be solved by converting the above linear equations to the Fourier space, yielding
\beq
    \tilde{\Psi}^{\pm}
    = -\frac{3\alpha \gamma^{+} \gamma^{-}}{\displaystyle 4\pi\left(\gamma^{+}+\gamma^{-}\right)}\frac{\displaystyle (\p\cdot\hat{\k})\p\cdot\left( \boldsymbol{F}(\tilde{\Psi}_{+}) - \boldsymbol{F}(\tilde{\Psi}_{-})\right) }
    {\displaystyle \sigma+k^2d_T^{\pm}+i\gamma^{\pm}(\hat{\k}\cdot\p)},
    \label{linear}
\eeq
where $\hat{\k} = \k/|\k|$ is a unit wave vector. In deriving Eq.~\eqref{linear}, we follow \cite{saintillan08a} to define the operator $
\boldsymbol{F}(\tilde{\Psi}^{\pm})=(\boldsymbol{I}-\hat{\k}\hat{\k})
\cdot\int_S\p(\p\cdot\hat{\k})\widetilde{\Psi}^{\pm}d S_p
$ and solve the fluid disturbance velocity in the Fourier space as
\beq   \tilde{\u}=\frac{i}{k}(\boldsymbol{I}-\hat{\k} \hat{\k}) \cdot \tilde{\bSig} \cdot \hat{\k},
\eeq
through which the two species interact with each other hydrodynamically.
Applying the operator of $\boldsymbol{F}(\tilde{\Psi}^{\pm})$ on both sides of Eq.~\eqref{linear} yields two coupled equations, from which we can solve the eigenvalue problem to obtain the dispersion relation
\begin{gather}
\gamma_{-}\left(2 a_+^{3}-\frac{4a_+}{3} +\left(a_+^{4}-a_+^{2}\right) \log \left(\frac{a_+-1}{a_++1}\right)\right)-
\gamma_{+} \left(2 a_-^{3}-\frac{4a_-}{3} +\left(a_-^{4}-a_-^{2}\right) \log \left(\frac{a_--1}{a_-+1}\right)\right) \nonumber \\
+ \frac{4ik\left(\gamma^{+}+\gamma^{-}\right)}{3 \alpha} =0
\label{dispersion}
\end{gather}
where $a_{\pm} = -i\left(\frac{\sigma+k^2d_T^{\pm}}{k\gamma^{\pm}} \right)$.

\begin{figure*}
 \begin{center}
  \includegraphics[width = 165mm]{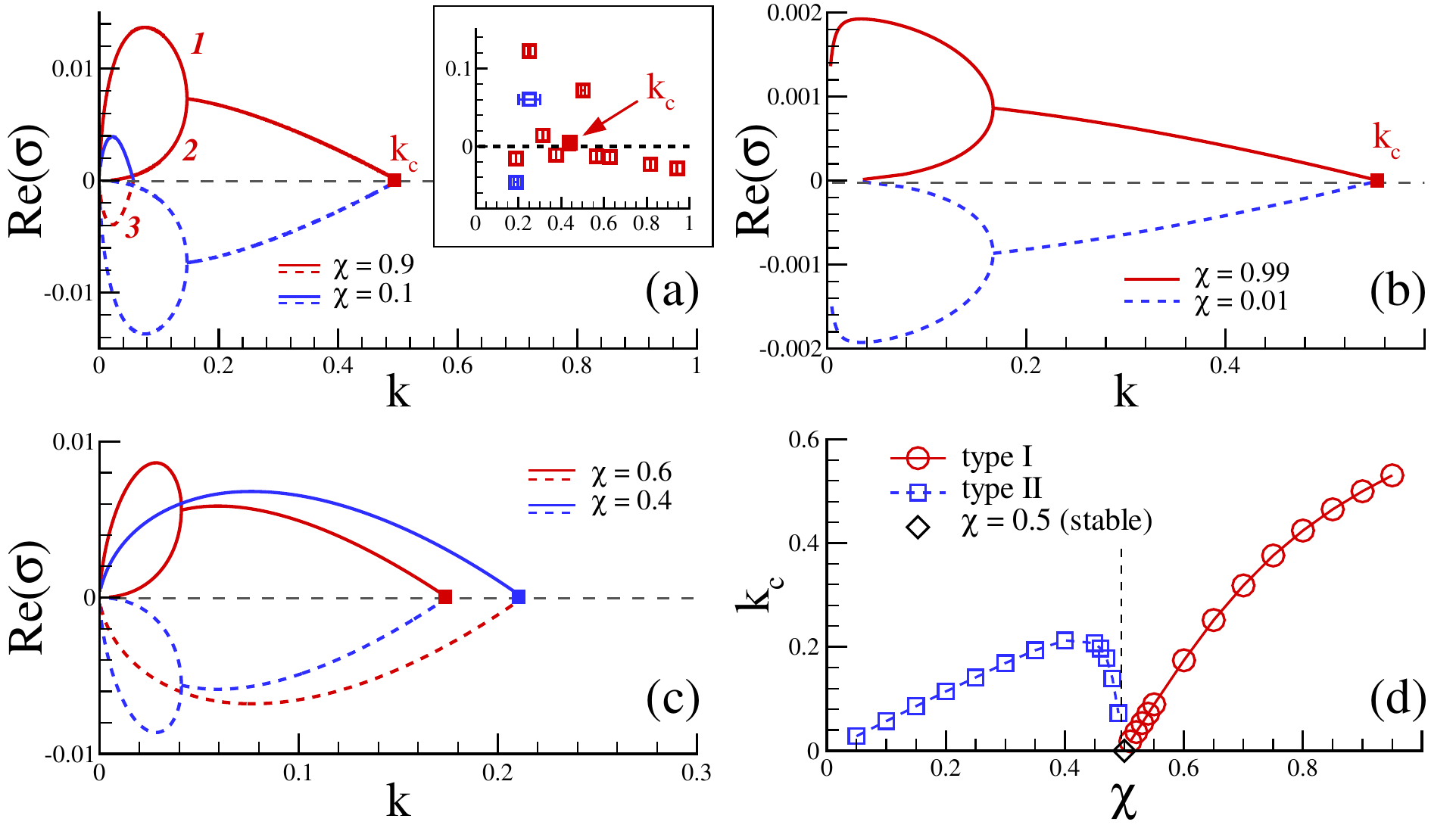}
\end{center}
\caption{Linear stability analysis of the kinetic model for activity-balanced binary mixtures. (a-c) Growth rate $\rm{Re}\left(\sigma\right)$ as a function of wavenumber $k$ at various choices of $\chi$. Inset of panel(a): $\rm{Re}\left(\sigma\right)-k$ curve obtained from particle simulation for the cases of $\chi = 0.9$ (red square) and $\chi = 0.1$ (blue square). (d) The critical wavenumber $k_c$ as a function of $\chi$ for the ``pusher-dominant'' (type I) and ``pusher-faster'' (type II) cases. The solid and dashed lines respectively represent the positive and negative growth rates. }
  \label{stab}
\end{figure*}
As shown in Fig.~\ref{stab}(a-c), we can numerically solve three solution branches for the real part of the growth rate $\rm{Re}\left(\sigma\right)$ from Eq.~\eqref{dispersion} at different values of $\chi$ when fixing $\alpha=-1$ and neglecting diffusion ($d^{\pm}_T= d^{\pm}_R =0$).  When pushers outnumber pullers (i.e., $\chi > 0.5$, marked by solid and dashed red lines) but with slower swimming speeds, the two solid branches on the top are unstable, while the third dashed branch with negative values is stable. Although there is no diffusion added, the system has intrinsic mechanisms to damp fluctuations at large $k$ for all cases considered, yielding a cut-off wavenumber $k_c$ beyond which all fluctuations become damped. Starting from $\chi$ close to 1, we observe the $\rm{Re}\left({\sigma}\right)$ decreasing as $\chi$ decreases, with a smaller and smaller $k_c$, until $\rm{Re}\left({\sigma}\right)$ drops to zero at $\chi = 0.5$, which is consistent with the prediction of stable symmetric binary mixture with effectively non-interacting particles by {B\'ardfalvy} {\it et al.} \cite{dora2020}.
As $\chi$ further decreases, we find the corresponding solutions marked by blue lines are exactly the mirror images with respect to the $k-$axis of those marked by red computed at $1-\chi$, meaning that all stable (resp. unstable) solutions are now flipped to become unstable (resp. stable). As a result, instabilities, while weak, may occur even when there are more pullers than pushers.

To make comparison with the kinetic model for a single species, we analyze the limiting behavior of our model as $\chi$ approaches 0 or 1. Note that at the extrema of $\chi=0 \text{ or } 1$, the activity-balanced condition from Eq. \eqref{stresscondition} requires that the particles have zero isolated swimming speed. This is unsurprising, as the only activity-balanced case within a single species suspension is that of a passive system. As demonstrated in Fig.~\ref{stab}(b), the magnitude of this finite wavelength instability decreases to zero in the limit as $\chi \rightarrow 0 \text{ or } 1$, indicating that the system is stable. At this limit, the critical wavenumber corresponding to maximum growth rate approaches zero while $k_c$ tends to 0.55, which hence suggests to recover the long-wavelength instability discussed in Refs. \cite{ramaswamy10,saintillan08a}.

The sign-flipping of the stable/unstable solutions effectively leads to non-monotonic variation of $k_c$ when plotted as a function of $\chi$ in Fig.~\ref{stab}(d). We observe that $k_c$ first increases with $\chi$ and then quickly drops to zero at $\chi = 0.5$ when symmetric binary suspensions are stable; when $\chi$ further increases, $k_c$ grows from zero again. Therefore, $\chi = 0.5$ defines a critical point of a first-order transition between the two types of hydrodynamic instabilities. Firstly, instability may arise because of a larger number of pushers, albeit moving slowly, which we referred to as the ``pusher-dominant'' (type I) cases. Although the shape of the unstable branches (solid red lines) resemble those obtained for the  single-species suspensions of pure pushers \cite{saintillan08a}, the maximum growth rates we obtained are one order of magnitude lower, which hence suggests much weaker instabilities. Also, these instabilities are of finite-wavelength with the maximum growth rate occurring at $k>0$, compared to the long-wave instabilities for pure pushers \cite{simha02,saintillan08a}. Secondly, the kinetic model suggests that at small values of $\chi$ when (slow) pullers outnumber (fast) pushers, instabilities may still occur, hence referred to as the ``pusher-faster'' (type II) cases that are marked by the blue color. However, we need to mention that in this scenario, instabilities are generally weaker than type I cases, typically with much smaller maximum growth rates and smaller $k_c$.

We have so far used a continuum model to successfully capture finite-wavenumber instability during the initial transient period. While intriguing, the occurrence of such instabilities poses a significant challenge in exploration of the system's longtime behavior using this kinetic model. The main difficulty is that, as the disturbance solutions keep growing, the fully developed unstable dynamics may feature significant fluctuations and inhomogeneity. Thus, in these scenarios, it will be invalid to convert the global activity-balance constraint to the local mean-stress-free condition by assuming uniform isotropy. As discussed in the next section, we instead turn to direct particle simulations to explore the late-time dynamics.

\section{Direct particle simulations}
\label{simu}

\subsection{Slender Body Model}

In this section, we reformulate and extend the slender body model by Saintillan and Shelley \cite{SS2012} to a binary suspension composed of $N$ total rods with center-of-mass (C.O.M.) position $\x$ and orientation along the unit vector $\p$. As sketched in Fig.~\ref{fig_schematic}, we use $\ell_h = \ell/2$ to denote the half-length of each rodlike particle and  $\{\dot{\x}_j,\dot{\p}_j\}_{j=1}^{N}$ to denote the translational and orientational velocities of each rod when subjected to both local actuation and the external net body force $\boldsymbol{F}^{e}$ and torque $\boldsymbol{T}^{e}$. In contrast with the slender body model by Saintillan and Shelley \cite{SS2012}, which models the rod's self-driven motion by imposing actuation stresses, we choose to impose the projected surface slip velocity $u_s^{\pm}(s)$, i.e. the projection of the true surface slip velocity onto the rod's center-line along the local arclength $s\in[-\ell_h,\ell_h]$.
We then denote the corresponding unknown force distribution along the particle center-line as $\boldsymbol{f}^e(s)$ and the induced disturbance velocity as $\u_\infty(s)$. The relation between particle motion and force distribution can then be calculated using slender-body theory as
\beq
    \dot{\x}+s \dot{\p}+u_{s}^{\pm}(s) \p-\u_{\infty}(s) = c(\boldsymbol{I}+\p\p) \cdot \boldsymbol{f}^{e}(s), \label{SBT}
\eeq
where the hydrodynamic force distribution $\boldsymbol{f}^e(s)$ satisfies the force- and torque-free conditions $\int_{-\ell_h}^{\ell_h} \boldsymbol{f}^{e}(s) d s = \boldsymbol{F}^{e}$ and $\int_{-\ell_h}^{\ell_h} s \p \times \boldsymbol{f}^{e}(s) d s = \boldsymbol{T}^{e}$ and $c = \log\left(2/r\right)/4\pi\mu$ is an effective drag coefficient. By taking the first and second moments of the orthonormal projection of \eqref{SBT} onto the parallel and perpendicular directions to $\p$, we can arrive at
\begin{gather}
    2\ell_h\dot{\x} = \boldsymbol{H}_\infty
    - H_{u s}^{\pm}\p  + c\left( \boldsymbol{I} + \p\p \right) \boldsymbol{F}^{e}, \label{xdotSBT}\\
    2 \ell_h^{3}\dot{\p} =   3(\boldsymbol{I}-\p \p) \cdot \boldsymbol{H}_\infty^s + 3 c \boldsymbol{T}^{e} \times \p. \label{pdotSBT}
\end{gather}
Here, we define a series of integral operators:
\begin{equation}
\boldsymbol{H}_{\infty}=\int_{-\ell_h}^{\ell_h} \u_{\infty}(s) d s, \quad
\boldsymbol{H}_{\infty}^s=\int_{-\ell_h}^{\ell_h} s\u_{\infty}(s) d s, \quad
H_{u s}^{\pm}=\int_{-\ell_h}^{\ell_h} u_s^{\pm}(s) d s, \quad H_{u s}^{s\pm}=\int_{-\ell_h}^{\ell_h} s u_s^{\pm}(s) d s,
\label{integ}
\end{equation}
which lead to the relation between the zeroth moment of slip velocity and the isolated particle velocity: $H_{us}^{\pm}=-\ell_h^2v^{\pm}=-\ell_h^2\gamma^{\pm}u_c$.

We perform simulations in the absence of any net external force and torque, i.e. $\boldsymbol{F}^{e}=\textbf{0}$ and $\boldsymbol{T}^{e}=\textbf{0}$, and ignore the effect of Brownian fluctuations. As well, we fully resolve the rod-rod collisions by utilizing a geometric constraint minimization technique to both restrict any particle penetrations and solve for the induced collision forces. The reader is referred to Ref \cite{yan2019} for more details on our collision algorithm.
The effect of the many-body interactions within Eq. \eqref{SBT}- \eqref{pdotSBT} are captured by the hydrodynamic disturbance velocity, $\u_\infty$, which can be calculated using the fundamental solution to the Stokes equations:
\begin{equation}\label{Green}
    \u_{\infty,\alpha}(\x)=\frac{1}{8 \pi \mu} \sum_{\substack{\beta=1 \\ \beta \neq \alpha}}^{N} \int_{-\ell_h}^{\ell_h} \boldsymbol{G}\left(\x ; \x_{\beta}+s_{\beta} \p_{\beta}\right) \cdot \boldsymbol{f}^e_{\beta}\left(s_{\beta}\right) d s_{\beta},
\end{equation}
where $\boldsymbol{G}(\x,\boldsymbol{x}')$ is the Greens function for Stokes flow known as the Stokeslet. Within this framework, instead of directly using the Stokeslet, we adopt the widely used Rotne-Prager-Yamakawa (RPY) tensor \cite{wajnryb13}, which guarantees symmetric-positive-definiteness whereas the Stokeslet does not. Physically, the positive-definiteness of the RPY tensor ensures that any non-zero force $\boldsymbol{f}^e$ will drive viscous dissipation into the surrounding fluid. Further, to accelerate the evaluation of the many-body interactions, we use a kernel-independent fast multipole method with triply periodic boundary conditions \cite{ying04,yan2018214} to efficiently reduce the computational cost of evaluating Eq.~\eqref{Green} from $O(N^2)$ to $O(N)$. After substituting Eq.~\eqref{xdotSBT} and \eqref{pdotSBT} into Eq.~\eqref{SBT}, we solve the linear system $\A \cdot \boldsymbol{f}^e = \boldsymbol{b}$ using iterative methods such as the generalized minimal residual (GMRES) method \cite{Saad86}. The matrix-free representation of $\boldsymbol{A}$ operating on the external force distribution, $\boldsymbol{f}^e$, and the corresponding $\boldsymbol{b}$ vector are defined as follows:
\begin{widetext}
\begin{gather}
    \A \cdot \boldsymbol{f}^{e} := \boldsymbol{f}^{e}+\frac{1}{c}\left(\boldsymbol{I}-\frac{1}{2} \p \p\right)\cdot \u_{\infty}(s)-\frac{1}{2 c \ell_h}\left(\boldsymbol{I}-\frac{1}{2} \p\p\right) \cdot \boldsymbol{H}_{\infty}-\frac{3 s}{2 c \ell_h^{3}}(\boldsymbol{I}-\p\p) \cdot \boldsymbol{H}_{\infty}^{s}, \label{Afe} \\
    \boldsymbol{b}:= -\frac{\p}{4c}\left(\frac{{H}_{u s}^{\pm}}{\ell_h}-2 u_{s}^{\pm}(s)\right) \label{b}.
\end{gather}
\end{widetext}

Next, we follow the previous section to derive the activity-balance condition for our particle simulations via analyzing the leading-order extra stress produced by all particle activity across the computation domain. Similar to our analysis of short-time dynamics, here we follow Batchelor's formulation \cite{batchelor70} to evaluate the effective extra stress exerted upon the ambient fluid due to motions of slender rods:
\beq
\bsig  =  - \frac{1}{V}\sum\limits_{i = 1}^N {\int_{ - {\ell _h}}^{{\ell _h}} s{{{{\f}}^e}\left(s\right){\p}} } ds = - \frac{1}{V}\sum\limits_{i = 1}^N {\sigma^{\pm}_0 \p\p}
%n{{\overline{ \sigma} }_0}{{\p\p}},
\eeq
where ${\sigma}^{\pm}_0 = \int_{-\ell_h}^{\ell_h} s\p\cdot\boldsymbol{f}^e d s $ represents the stresslet on a single pusher or puller. From Eq.~\eqref{SBT}, it is straightforward to solve for $\sigma_0^{\pm}$ as
\begin{equation}
    \sigma_0^{\pm} =  -\frac{1}{2c}H_{us}^{s\pm} + \frac{1}{2c}\p\cdot\boldsymbol{H}_{\infty}^{s}.
        \label{sigmaApprox}
\end{equation}
Then, we can define the global mean stresslet, $\overline{\sigma}_0$ (the overline denotes a spatial average), as
\beq
{{\overline{ \sigma }}_0} = \frac{1}{N}\sum\limits_{i = 1}^{{N^ + }} {\sigma _0^ + }  + \frac{1}{N}\sum\limits_{i = 1}^{{N^ - }} {\sigma _0^ - }  =  - \frac{1}{{2c}}\left( {\chi H_{us}^{s + } + \left( {1 - \chi } \right)H_{us}^{s - }} \right) + \frac{1}{{2cN}}\sum\limits_{i = 1}^N {\p \cdot \boldsymbol{H}_{\infty}^{s}}.
\label{globalmean}
\eeq
Note that when performing simulations for dilute suspensions,  the extra stress induced by the individual particle's self-swimming motion will dominate over that induced by hydrodynamically mediated many-body interactions (see our simulation results for validation of this assertion). In other words, the isolated particle stresslet will be dominant causing $||\boldsymbol{H}_{\infty}^{s}|| \ll |H_{us}^{s\pm}|$. Hence, we can derive the activity-balance condition by enforcing the leading-order term on the right-hand-side of Eq.~\eqref{globalmean} to be zero
\beq
H_{us}^{s+}\chi + H_{us}^{s-}\left(1-\chi\right) = 0,
\label{stressbalancecondition}
\eeq
where $\chi$ is the relative number fraction defined above. It is worthwhile to mention that the above condition is derived solely from the global mean stress without the need to construct ``local'' extra stresses for discrete particles.

All  simulation  presented  herein are performed in a cubic periodic box of length $L=10 \mu m$ with pushers and pullers of the equal length $\ell=1 \mu m$ and aspect ratio $A = \ell/b = 10$. For simplicity, we choose to represent all particle slip velocities using the simple Heaviside step function $u^{\pm}_s(s) = - v^{\pm} \left[1 \mp \rm{sgn}(s)\right],~v^{\pm}>0$. Hence, when taking the integral operator as $H_{us}^{s\pm} = \pm\ell_h^2v^{\pm} = \pm\ell_h^2\gamma^{\pm}u_c$ defined in Eq.~\eqref{integ}, the activity-balance condition in \eqref{globalmean} can be simplified to
\begin{equation} \label{stressbalanced}
    \frac{\gamma^+}{\gamma^-} = \frac{1-\chi}{\chi},
\end{equation}
i.e., the ratio of each species' isolated swimming speed be inversely proportional to the ratio of their number fraction, which recover the condition derived in \eqref{stresscondition}. Notice that for a given $\chi$, Eq.~\eqref{stressbalanced} only constrains the ratio of each species' isolated speed. To then obtain a value for $\gamma^{\pm}$ from this condition, we constrain the mean isolated swimming speed of each species using $\beta = \frac{\gamma^+ + \gamma^-}{2}$. After validating our selection of $\beta$, we systematically explore the simulation parameter space by varying the global number density $n$ and relative number fraction $\chi$.

%Note, this stress balanced condition is the same in form as the one derived within our continuum formulation Eq. \eqref{stresscondition}; the difference between the two is that our continuum formulation requires additional restrictions on the concentration and orientation distributions for the stress balanced condition to be satisfied. As a result, the our particle simulations satisfy the stress balanced condition only on a global scale and not on a smaller continuum scale since local variations within the concentration and orientation distribution are allowed to occur.

%In addition to the slender body formulation, an additional aspect of our particle model is our collision resolution algorithm which, put simply, applies a geometric constant to the equations of motion such that at the end of each time step no overlap exists between particles. This geometric constraint can be reformulated into a linear complementarity problem which reduces the determination of the collision force to a global optimization problem. Because our current simulations are within the dilute regime, the effect of collisions plays a small role in comparison to hydrodynamics. As a result,  Though, it is worthy to note that we currently solve the the motion induced by each collision force by a block diagonal mobility problem which neglects the effects of many body hydrodynamics.

\begin{figure*}
 \begin{center}
  \includegraphics[width = 168mm]{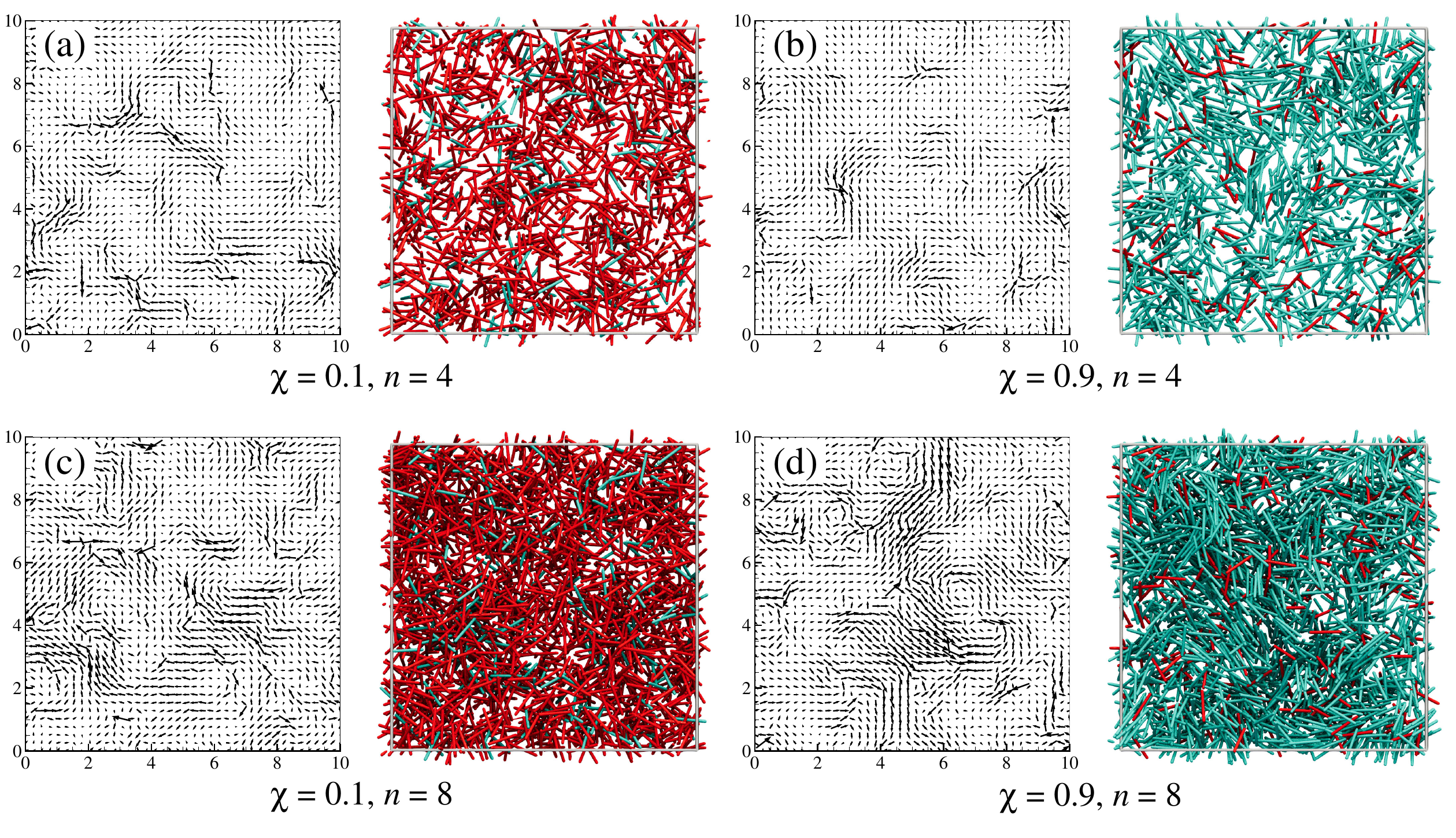}
 \end{center}
   \caption{Snapshots of typical flow fields and rod distributions obtained from particle simulations in a cubic box of size $10 \times10 \times10$ at $t = 100s$ for number density $n = 4, 8$ (equivalent to total particle number $N = 4000, 8000$) and pusher number fraction $\chi = 0.1, 0.9$. The pushers and pullers are marked by green and red colors, respectively. The planar vector plot on the left corresponds to the 3D fluid velocity field in each panel when projected on the XY-plane. On the right, rendered images show the rod distribution within a sub-section of dimensions $10 \times10 \times 3$. }
 \label{flow}
\end{figure*}

\subsection{Simulation Results}

Consistent with the mean-field analysis, all simulations presented herein start from an initially isotropic configuration where all rods are homogeneously distributed with random orientations. These simulations are ran over a long period ($t > 50s$) until the system reaches quasi-steady states.

Different mixed systems can be classified as ``stable'' or ``unstable'' with those long-time simulations after reaching steady states.
However, the linear instability spectrum $\sigma(k)$ Eq.~\eqref{plane-wave} calculated by the continuum model is only for the very short initial period where the deviation from isotropic distribution function is small and linearizable.
Direct measurement of the growth rate $\sigma(k)$ from particle simulations is very challenging and rarely performed in the past, mostly because the deviation from isotropic function is a very weak signal, usually hidden in various sources of noises and numerical errors in discrete simulations.
To extract the weak signal, we perform an ensemble average over a large number of short simulation runs.
The initial configuration of each simulation has identical center-of-mass locations but randomly initialized orientations.
For each case, we compute the distribution function $\Psi(\bm{k},t)$ by a non-uniform fast-Fourier transform method \cite{barnett19}.
Each of these simulations is ran for $4s$ during which each rod swims at most 4 body-lengths to induce sufficient amount of hydrodynamic inter-particle interactions but short enough to ensure that we remain within the initial linear-instability regime.
Then we average $\Psi(\k,t)$ over 400 simulations.
Therefore, we can average out the noise induced by orientations and focus on the instability spectrum generated by isotropic distribution function.
We then fit the linear growth factor $\sigma(k)$ for each k with the averaged $\langle\Psi(\bm{k},t)\rangle$, the results for which are shown in the inset of Fig.~\ref{stab}(a).
With this method, we do qualitatively captured the weak instability predicted by the kinetic model for both I and II type when selecting $\chi = 0.1$ and $0.9$.
Especially for the pusher-dominant case at $\chi = 0.9$ (red open squares), we found that the cutoff wavenumber $k_c\approx0.45$ with simulations, close to the continuum model prediction $k_c = 0.5$; for the pusher-faster case at $\chi = 0.1$, we also obtained a positive growth rate around $k = 0.25$. We need to mention that the main challenge here is that in the long-wavelength limit ($k\to 0$), the discrete $k$ vectors are very sparse, limited by the simulation box size and the noise is still significant even after averaging over 400 simulations.

%{\bf To make connections to the stability analysis, we first examine the short-time transient for the first $XX s$ during which each rod approximately swims over one bodylength to generate a significant amount of rod-rod interactions.
%results in Section \ref{simu}, for a given number density $n$, $k_c$ in fact defines a critical dimensionless domain size $L/\ell^p_c$ ($\ell^p_c$ represents the length scale selected in the particle simulation) beyond which instability occurs \cite{HS2010,SS2012}, i.e.,
% ${L}/{\ell^p_c} \geq {2\pi}/{k_c}$.
%Here we follow Ref. \cite{saintillan07} to use the half rod-length $\ell_h$ to define $\ell^p_c = \ell_h/\nu$ with $\nu = N \left(\ell_h/L\right)^3$ an effective volume fraction. When taking the cut-off value $k_c \approx 0.5$ at $\chi = 0.9$ in panel(a), we can predict that there must be at least a total of $N \approx 5000$ rods in a computation domain of size $L/\ell = L/2\ell_h = 10$ for instabilities to occur, which is in agreement with our simulation results. For example, as shown in Fig.~\ref{corr}, the velocity and orientation-order correlations functions at $n=6$ ($N=6000$, blue lines) and $8$ ($N=8000$, red lines) are close to each other, and show significantly higher correlations than $n=4$ (i.e., $N=4000$) case. Nevertheless, so far we haven't capture clear longtime correlated motions for the puller-dominant cases (i.e., the type II instability) directly from particle simulations. }

Next, we collect statistic measurements at late times ($t > 50 s$). As shown in Fig.~\ref{flow}(a-d), the snapshots in each panel are taken for typical 2D flow fields on a XY-plane (on the left), and the corresponding particle distributions within a thin sub-section of size $10\times 10\times 3$ (on the right). When both $n$ and $\chi$ are low, the rods remain homogeneously distributed across the domain without forming clear structures. Even for large $n$, so long as $\chi$ remains low, the rods appear randomly oriented (see movie S1); on the other hand, as $\chi$ is increased, we clearly see spatial inhomogeneity characterized by the formation of local clusters and alignment of rods (see movie S2). As shown in panel(d) for the case of large $n$ ($n = 8$ or equivalently $N = 8000$) and large $\chi$ ($\chi=0.9$) when the (slow) pushers significantly outnumber the (fast) pullers, we can clearly observe large-scale correlated motions and the formation of local clusters accompanied by small but noticeable density fluctuations. In the meantime, the corresponding flow fields appear to be reminiscent of jet-like and vortical structures.

% It appears that in most cases, the rods remain homogeneously distributed across the domain without forming clear local structures (see movie S1), especially when both $n$ and $\chi$ are low. In the meantime, the corresponding flow fields induced by rods' motions only feature small-scale fluctuations. However, as shown in panel(d) for the case at a large $n$ ($n = 8$ or equivalently $N = 8000$) and $\chi$ ($\chi=0.9$, i.e., when (slow) pushers significantly outnumber (fast) pullers), we can clearly observe large-scale correlated motions (see movie S2). In the meantime, the corresponding flow fields appear to be reminiscent of jet-like and vortical structures, accompanying by small but noticeable density fluctuations.

\begin{figure}[t]
  \centering
  \includegraphics[width=1\textwidth]{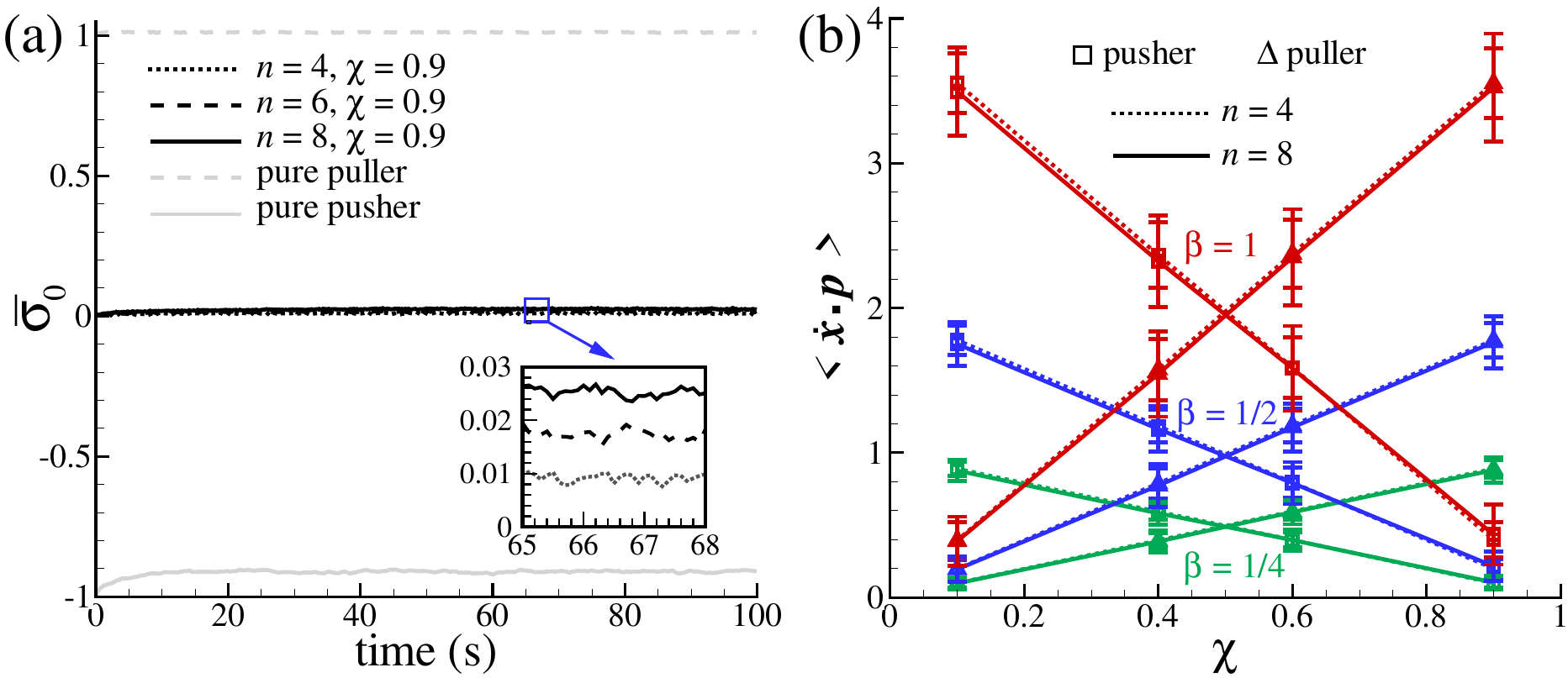}
  \caption{Global means of the late-time dynamics of particle simulations. (a) Mean stresslet $\overline{\sigma}_0$ (scaled by $\mu u_c\ell^2$) as a function of time. (b) Mean projected C.O.M. speed $\langle\dot{\x}\cdot\p\rangle$ (scaled by $u_c$) as a function of $\chi$ with standard deviation error bars. Three mean isolated swimming speeds, $\beta = \frac{\gamma^+ + \gamma^-}{2}$ with $\beta = 1/2, 1, 2$, are chosen. In (a) and (b), we also vary number density $n = 4, 6, 8$, which are equivalent to total particle number $N = 4000, 6000, 8000$.}
\label{cross}
\end{figure}

To gain a more quantitative understanding, we begin with measuring global quantities via spatial and ensemble (denoted by an angle bracket ``$\langle \rangle$'') averaging. As shown in Fig.~\ref{cross}(a) for typical cases of large $\chi$ ($\chi = 0.9$) at different number densities, even at a quasi steady state, the global mean stresslet $\overline{\sigma}_0$ of our activity-balanced simulations always remains close to zero with only small deviations (see the zoomed-in inset), which is in contrast to the pure pusher/puller cases (grey lines). This is consistent with our assumption that the system is dilute and that the activity induced stress is dominant. In panel(b), we examine the (dimensionless) mean and standard deviation of the C.O.M. speed when projected along the rod's orientation direction, i.e., $\langle \dot{\x} \cdot \p \rangle$, which are calculated separately for the pushers and pullers when varying $\chi$ and $n$. It is clearly observed that for all cases considered here, the global mean values only slightly deviate from the isolated single-particle speeds with little fluctuations, instead of exhibiting large velocity standard deviations and enhanced mean speeds that are typically seen in pure pusher suspensions \cite{SS2012}. Varying $\beta$ had little impact on the projected C.O.M speed nor any of the results presented herein; therefore, without loss of generality, we present all results for $\beta=1$. We have also measured the  orientational order parameter $S=\sqrt{\frac{3}{2}\left\langle \p\p - \frac{\boldsymbol{I}}{3} \right\rangle : \left\langle \p\p - \frac{\boldsymbol{I}}{3} \right\rangle}$ and found that the ensemble average value remains near zero, i.e., $\langle S  \rangle = 10^{-3} \sim 10^{-2}$, for all simulations.

\begin{figure}[t]
  \centering
  \includegraphics[width=1\textwidth]{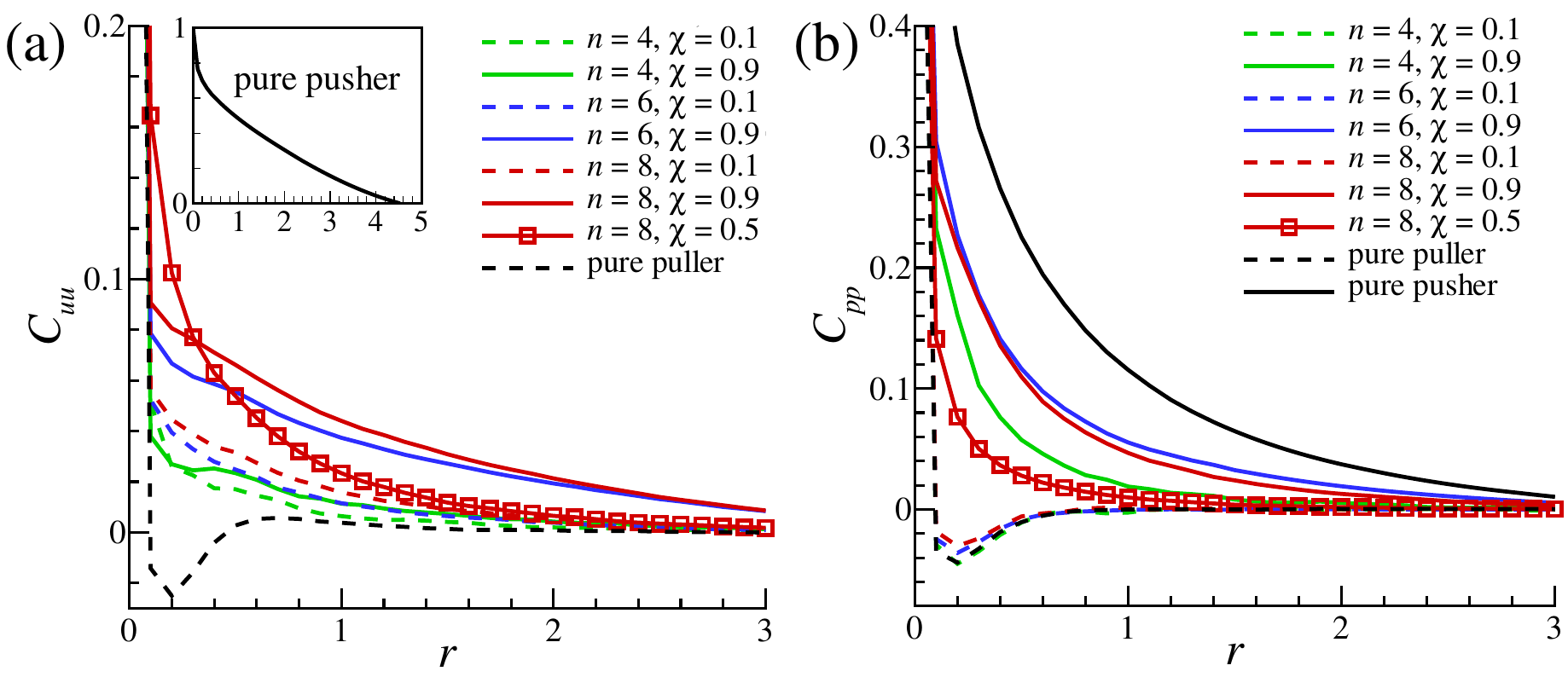}
  \caption{Two-point correlation functions for rods' (a) C.O.M. velocity ($C_{uu}$) and (b) orientation ($C_{pp}$) fields, with $r$ the dimensionless distance scaled by $\ell$. The black color marks the activity-unbalanced cases of pure pushers and pure pullers. Inset in (a): $C_{uu}$ as a function of $r$ for the pure pusher case at $n = 8$.}
\label{corr}
\end{figure}

Next, we measure two-point radial correlation functions to reveal more details about the spatial variations of the unstable dynamics at quasi-steady states. First of all, we define two correlation functions $C_{u u}$ and $C_{p p}$ for the particle's C.O.M. velocity $\dot{\x}$ and orientation $\p$, respectively,
\beq
    C_{u u}(r) = \frac{\langle \dot{\x}(0) \cdot \dot{\x}(r)\rangle} { \langle |\dot{\x}(0)| \rangle^2}, \quad
    C_{p p}(r) = \frac{\langle \p(0) \cdot \p(r)\rangle} {\langle |\p(0)| \rangle^2}
\eeq
where $r$ is the dimensionless radius (scaled by $\ell$) of the virtual spherical shell centered within each particle during sampling. As shown by some typical cases in Fig.~\ref{corr}, when choosing $n=4,6, \text{ and }8$, we observe that the correlation functions often exhibit monotonic decays towards zero for large $r$. Therefore, to appropriately define the correlation lengths, instead of using the $r$ location corresponding to either the first minima or the first zero (i.e., cut-off length), we define the correlation lengths when the function becomes near zero
\begin{equation}
   r^c:= \left\{r^c\in \left[0,\frac{L}{2\ell}\right]: C(r^c) = 0.01 \leq C(s) \quad \forall  s\in\left[0,r^c\right]\right\}
\end{equation}
where the maximum attainable correlation length is limited by the half-length of our simulation domain. To make fair comparisons, we added two stress-unbalanced cases for pure pushers and pure pullers (black solid and dashed lines), where the hydrodynamic correlations are well understood.

\begin{figure*}
 \begin{center}
  \includegraphics[width = 168mm]{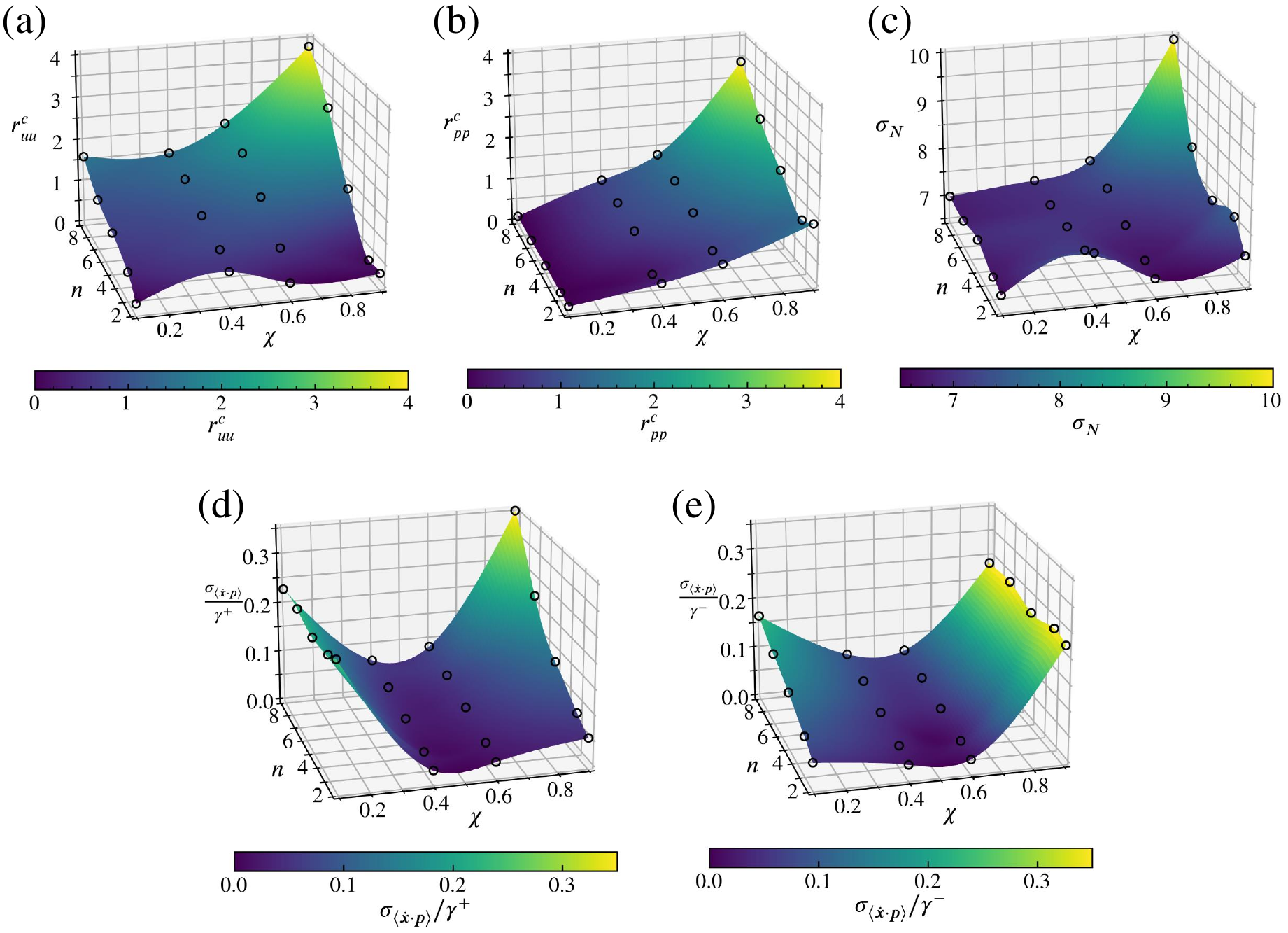}
 \end{center}
   \caption{Phase diagrams of statistic measurements from particle simulations in terms of $n \in [1,8]$ and $\chi \in [0.1,0.9]$: (a) rod velocity correlation length $r^c_{uu}$; (b) rod orientation correlation length $r^c_{pp}$; (c) Standard deviation of rod density distribution $\sigma_N$. (d-e) Normalized standard deviation of the projected C.O.M. speed, $\langle\dot{\x}\cdot\p\rangle$ for each species. The black open circles represent the calculated data.}
 \label{phase}
\end{figure*}

As shown in Fig.~\ref{corr}(a), $C_{uu}$ suggests that the activity-balanced cases have weaker velocity correlations with correlation length $r^c_{uu} = 2 \sim 3$, compared to the pure pusher case in the inset with $r^c_{uu} = 4.5$, approximately the half-domain size. But, even for those puller-dominant cases at $\chi<0.5$, their velocity correlations appear to be much stronger than the pure puller case (black dashed line) where $C_{uu}$ drops below zero at a small $r$, indicating weak anti-correlation. In panel(b), the $C_{pp}-r$ curves extracted from the orientation fields exhibit much more significant correlations and cleaner trends than the $C_{uu}-r$ curves in panel(a). Interestingly, we observe that when approximately $\chi > 0.5$ (solid lines), $C_{pp}$ appears to exhibit a strictly positive and monotonic decay, reminiscent of the pure pusher case (black solid line). Meanwhile, at $\chi < 0.5$ (dashed lines), $C_{pp}$ varies non-monotonically and can first drop below zero like the pure puller case to show slight anti-correlations. In addition, the orientation correlation length of $r^c_{pp}$ for the stress-balanced cases are in fact, comparable to the stress-unbalanced ones. For large values of $n$ and $\chi$, e.g. when $\chi = 0.9$ and $n=8$, we obtain $r^c_{pp} \approx 2.4$ compared to $r^c_{pp} \approx 3.2$ measured for the pure pusher case. We collect the measured values of $r^c_{uu}$ and $r^c_{pp}$, and plot them against $\left(n,\chi\right)$ in Fig.~\ref{phase}(a) and (b), respectively. There are ``hot spots'' occurring at large $n$ and $\chi$ occurring where $r^c_{uu}$ and $r^c_{pp}$ show a strong dependence on both parameters and indicate significant correlated motions when pusher-pusher interactions are strengthened. In the other parameter regimes that are further away from the hot spots, $r^c_{uu}$ appear to vary linearly with $n$; while $r^c_{pp}$ exhibits approximate linear dependence on $\chi$.

In addition, we are able to quantify the density and velocity fluctuations in the stress-balanced binary suspensions. Here we follow Ref~\cite{SS2012} by taking a cubic sampling box of volume $\tilde{V}$ centered around each rod such that the expected number of particles in the box is selected as \beq
\langle \tilde{N} \rangle = \frac{\tilde{V}}{V}N = 50.
\eeq
Any density fluctuations are captured by taking the ensemble averaged standard deviation of particles within the search box, which we denote by $\sigma_N$. Similarly, the velocity fluctuations of each species are captured by the ensemble averaged standard deviation of the projected C.O.M. speed, $\langle\dot{\x}\cdot\p\rangle$, which we normalize by isolated swimming speed. For spatially homogeneous suspensions, their density distribution can be approximated by a Poisson distribution with a standard deviation of $\sigma_N=\sqrt{\langle \tilde{N}\rangle} \approx 7$. Nevertheless, as shown in Fig.~\ref{phase}(c) for the phase diagram of $\sigma_N$, the measured value may exceed that of a Poison distribution in the parameter regime of large $n$ and $\chi$. Again, we find these values are comparable with those of stress-unbalanced cases. For example, $\sigma_N\approx9.93$ is measured at $n=8$ and $\chi=0.9$, which is close to $\sigma_N\approx10.3$ obtained for the pure pusher case. Furthermore, we present the results for the measured velocity fluctuations in Fig.~\ref{corr}(d, e). Notice that the dominant, slow-moving species exhibits strong excitation by the less-prevalent, fast-moving species and that this excitation becomes stronger with increased number density. For example, at $\chi=0.9$ within Fig.~\ref{corr}(d), the standard deviation of the pushers increases by a factor of seven as the number density is increased from 1 to 8. On the other hand, little to no excitation is seen within the less-prevalent, fast-moving species. For example, at $\chi=0.9$ within Fig.~\ref{corr}(e), the projected center of mass standard deviation of the pushers remains fairly constant with $n$. Interestingly, the right hand sides of Fig.~\ref{corr}(a) and Fig.~\ref{corr}(d) display similar trends, as do the left hand sides of Fig.~\ref{corr}(a) and Fig.~\ref{corr}(e). This trend indicates that increased velocity correlation length is associated with an increase in velocity fluctuations and the ability of the less-prevalent, fast-moving species to excite the more-prevent, slow moving one.

\section{Conclusion and Discussion}\label{sec:conclusion}
To summarize, we have built a computation framework to study hydrodynamic instabilities and collective dynamics arising from dilute, activity-balanced pusher-puller binary suspensions, which produces near zero mean extra stress. Combining a mean-field kinetic model and large-scale discrete particle simulations, we have successfully demonstrated that in such binary systems, mixing pushers and pullers may lead to much richer dynamics and behaviors than simple neutralization processes as a result of the cancellation of stresslets from both species. Even without producing mean extra stresses, finite-wavenumber hydrodynamic instabilities can still grow from a uniform isotropic state and may eventually lead to well-correlated collective motions and non-trivial field fluctuations. These findings demonstrate that non-zero mean extensile stress is not necessarily a sufficient indicator of collective motions; this result differs from the classical single-species systems where net extensile stress is considered to be a clear indicator of collective motions.

We expect this work to initiate new computational and experimental studies of measuring, characterizing, and predicting the complex dynamics in multispecies-multicomponent active systems. Especially considering the high computation costs of large-scale discrete particle simulations, of particular interest will be developing more accurate continuum models that can fully resolve the late-time nonlinear dynamics arising from non-equilibrium, inhomogeneous mixtures. As discussed in Section \ref{sec:kinetic}, one key step in mean-field models like ours is constructing a local total extra stress (i.e., $\bSig$ in Eq.~\eqref{totstress}) to drive active flows and account for inhomogeneity. Besides formulating the concentration-dependent dipolar stresses (note that $\D$ scales with the local concentration $c$), we need to evaluate $\bSig$ based on a certain local fractional number, e.g., $\chi^{\pm}(\x) = \frac{c^{\pm}}{c^+ + c^-}$ , instead of simply treating $\chi$ as a constant global mean. Alternatively, one may choose the kinetic model developed by \v{S}kult\'ety {\it et al.} \cite{Viktor20} that avoids to use mean-field approaches for modeling the many-body interactions via directly constructing extra stresses. In their method, microparticles are treated as simple point dipoles. Then the induced hydrodynamic effects are evaluated by summing individual dipolar stresses via regularized Stokeslets. More computation studies of binary mixtures using continuum models, together with detailed comparisons with discrete particle simulations, will be presented in the future.

\begin{acknowledgements}
This work is supported by NSF Grant No. CAREER-1943759, and has used the computation resources from the Flatiron Institute and MSU's High Performance Computing Center. T.G. also acknowledges fruitful discussions with M. Shelley and D. Saintillan regarding slender-body model for particle simulations.
\end{acknowledgements}

\bibliography{draft}

%apsrev4-2.bst 2019-01-14 (MD) hand-edited version of apsrev4-1.bst
%Control: key (0)
%Control: author (72) initials jnrlst
%Control: editor formatted (1) identically to author
%Control: production of article title (-1) disabled
%Control: page (0) single
%Control: year (1) truncated
%Control: production of eprint (0) enabled
\begin{thebibliography}{26}%
\makeatletter
\providecommand \@ifxundefined [1]{%
 \@ifx{#1\undefined}
}%
\providecommand \@ifnum [1]{%
 \ifnum #1\expandafter \@firstoftwo
 \else \expandafter \@secondoftwo
 \fi
}%
\providecommand \@ifx [1]{%
 \ifx #1\expandafter \@firstoftwo
 \else \expandafter \@secondoftwo
 \fi
}%
\providecommand \natexlab [1]{#1}%
\providecommand \enquote  [1]{``#1''}%
\providecommand \bibnamefont  [1]{#1}%
\providecommand \bibfnamefont [1]{#1}%
\providecommand \citenamefont [1]{#1}%
\providecommand \href@noop [0]{\@secondoftwo}%
\providecommand \href [0]{\begingroup \@sanitize@url \@href}%
\providecommand \@href[1]{\@@startlink{#1}\@@href}%
\providecommand \@@href[1]{\endgroup#1\@@endlink}%
\providecommand \@sanitize@url [0]{\catcode `\\12\catcode `\$12\catcode
  `\&12\catcode `\#12\catcode `\^12\catcode `\_12\catcode `\%12\relax}%
\providecommand \@@startlink[1]{}%
\providecommand \@@endlink[0]{}%
\providecommand \url  [0]{\begingroup\@sanitize@url \@url }%
\providecommand \@url [1]{\endgroup\@href {#1}{\urlprefix }}%
\providecommand \urlprefix  [0]{URL }%
\providecommand \Eprint [0]{\href }%
\providecommand \doibase [0]{https://doi.org/}%
\providecommand \selectlanguage [0]{\@gobble}%
\providecommand \bibinfo  [0]{\@secondoftwo}%
\providecommand \bibfield  [0]{\@secondoftwo}%
\providecommand \translation [1]{[#1]}%
\providecommand \BibitemOpen [0]{}%
\providecommand \bibitemStop [0]{}%
\providecommand \bibitemNoStop [0]{.\EOS\space}%
\providecommand \EOS [0]{\spacefactor3000\relax}%
\providecommand \BibitemShut  [1]{\csname bibitem#1\endcsname}%
\let\auto@bib@innerbib\@empty
%</preamble>
\bibitem [{\citenamefont {Ramaswamy}(2010)}]{ramaswamy10}%
  \BibitemOpen
  \bibfield  {author} {\bibinfo {author} {\bibfnamefont {S.}~\bibnamefont
  {Ramaswamy}},\ }\href@noop {} {\bibfield  {journal} {\bibinfo  {journal}
  {Ann. Rev. Cond. Matt. Phys.}\ }\textbf {\bibinfo {volume} {1}},\ \bibinfo
  {pages} {323} (\bibinfo {year} {2010})}\BibitemShut {NoStop}%
\bibitem [{\citenamefont {Shelley}(2016)}]{Shelley2016}%
  \BibitemOpen
  \bibfield  {author} {\bibinfo {author} {\bibfnamefont {M.}~\bibnamefont
  {Shelley}},\ }\href@noop {} {\bibfield  {journal} {\bibinfo  {journal} {Annu.
  Rev. Fluid Mech.}\ }\textbf {\bibinfo {volume} {48}},\ \bibinfo {pages} {487}
  (\bibinfo {year} {2016})}\BibitemShut {NoStop}%
\bibitem [{\citenamefont {Ishikawa}\ \emph {et~al.}(2006)\citenamefont
  {Ishikawa}, \citenamefont {Simmonds},\ and\ \citenamefont
  {Pedley}}]{Ishikawa06}%
  \BibitemOpen
  \bibfield  {author} {\bibinfo {author} {\bibfnamefont {T.}~\bibnamefont
  {Ishikawa}}, \bibinfo {author} {\bibfnamefont {M.}~\bibnamefont {Simmonds}},\
  and\ \bibinfo {author} {\bibfnamefont {T.}~\bibnamefont {Pedley}},\
  }\href@noop {} {\bibfield  {journal} {\bibinfo  {journal} {J. Fluid Mech.}\
  }\textbf {\bibinfo {volume} {568}},\ \bibinfo {pages} {119} (\bibinfo {year}
  {2006})}\BibitemShut {NoStop}%
\bibitem [{\citenamefont {Saintillan}\ and\ \citenamefont
  {Shelley}(2008)}]{saintillan08a}%
  \BibitemOpen
  \bibfield  {author} {\bibinfo {author} {\bibfnamefont {D.}~\bibnamefont
  {Saintillan}}\ and\ \bibinfo {author} {\bibfnamefont {M.}~\bibnamefont
  {Shelley}},\ }\href@noop {} {\bibfield  {journal} {\bibinfo  {journal} {Phys.
  Fluids}\ }\textbf {\bibinfo {volume} {20}},\ \bibinfo {pages} {123304}
  (\bibinfo {year} {2008})}\BibitemShut {NoStop}%
\bibitem [{\citenamefont {Ben-Jacob}\ \emph {et~al.}(2016)\citenamefont
  {Ben-Jacob}, \citenamefont {Finkelshtein}, \citenamefont {Ariel},\ and\
  \citenamefont {Ingham}}]{BenJacob16}%
  \BibitemOpen
  \bibfield  {author} {\bibinfo {author} {\bibfnamefont {E.}~\bibnamefont
  {Ben-Jacob}}, \bibinfo {author} {\bibfnamefont {A.}~\bibnamefont
  {Finkelshtein}}, \bibinfo {author} {\bibfnamefont {G.}~\bibnamefont
  {Ariel}},\ and\ \bibinfo {author} {\bibfnamefont {C.}~\bibnamefont
  {Ingham}},\ }\href@noop {} {\bibfield  {journal} {\bibinfo  {journal} {Trends
  Microbiol.}\ }\textbf {\bibinfo {volume} {24}},\ \bibinfo {pages} {257}
  (\bibinfo {year} {2016})}\BibitemShut {NoStop}%
\bibitem [{\citenamefont {Wittmann}\ \emph {et~al.}(2018)\citenamefont
  {Wittmann}, \citenamefont {Brader}, \citenamefont {Sharma},\ and\
  \citenamefont {Marconi}}]{wittmann_effective_2018}%
  \BibitemOpen
  \bibfield  {author} {\bibinfo {author} {\bibfnamefont {R.}~\bibnamefont
  {Wittmann}}, \bibinfo {author} {\bibfnamefont {J.}~\bibnamefont {Brader}},
  \bibinfo {author} {\bibfnamefont {A.}~\bibnamefont {Sharma}},\ and\ \bibinfo
  {author} {\bibfnamefont {U.}~\bibnamefont {Marconi}},\ }\href@noop {}
  {\bibfield  {journal} {\bibinfo  {journal} {Phys. Rev. E}\ }\textbf {\bibinfo
  {volume} {97}},\ \bibinfo {pages} {012601} (\bibinfo {year}
  {2018})}\BibitemShut {NoStop}%
\bibitem [{\citenamefont {Maloney}\ \emph {et~al.}(2020)\citenamefont
  {Maloney}, \citenamefont {Liao}, \citenamefont {Klapp},\ and\ \citenamefont
  {Hall}}]{maloney_clustering_2020}%
  \BibitemOpen
  \bibfield  {author} {\bibinfo {author} {\bibfnamefont {R.}~\bibnamefont
  {Maloney}}, \bibinfo {author} {\bibfnamefont {G.}~\bibnamefont {Liao}},
  \bibinfo {author} {\bibfnamefont {S.}~\bibnamefont {Klapp}},\ and\ \bibinfo
  {author} {\bibfnamefont {C.}~\bibnamefont {Hall}},\ }\href@noop {} {\bibfield
   {journal} {\bibinfo  {journal} {Soft Matter}\ }\textbf {\bibinfo {volume}
  {16}},\ \bibinfo {pages} {3779} (\bibinfo {year} {2020})}\BibitemShut
  {NoStop}%
\bibitem [{\citenamefont {Stenhammar}\ \emph {et~al.}(2015)\citenamefont
  {Stenhammar}, \citenamefont {Wittkowski}, \citenamefont {Marenduzzo},\ and\
  \citenamefont {Cates}}]{stenhammar_activity-induced_2015}%
  \BibitemOpen
  \bibfield  {author} {\bibinfo {author} {\bibfnamefont {J.}~\bibnamefont
  {Stenhammar}}, \bibinfo {author} {\bibfnamefont {R.}~\bibnamefont
  {Wittkowski}}, \bibinfo {author} {\bibfnamefont {D.}~\bibnamefont
  {Marenduzzo}},\ and\ \bibinfo {author} {\bibfnamefont {M.}~\bibnamefont
  {Cates}},\ }\href@noop {} {\bibfield  {journal} {\bibinfo  {journal} {Phys.
  Rev. Lett.}\ }\textbf {\bibinfo {volume} {114}},\ \bibinfo {pages} {018301}
  (\bibinfo {year} {2015})}\BibitemShut {NoStop}%
\bibitem [{\citenamefont {Kolb}\ and\ \citenamefont
  {Klotsa}(2020)}]{kolb_active_2020}%
  \BibitemOpen
  \bibfield  {author} {\bibinfo {author} {\bibfnamefont {T.}~\bibnamefont
  {Kolb}}\ and\ \bibinfo {author} {\bibfnamefont {D.}~\bibnamefont {Klotsa}},\
  }\href@noop {} {\bibfield  {journal} {\bibinfo  {journal} {Soft Matter}\
  }\textbf {\bibinfo {volume} {16}},\ \bibinfo {pages} {1967} (\bibinfo {year}
  {2020})}\BibitemShut {NoStop}%
\bibitem [{\citenamefont {Takatori}\ and\ \citenamefont
  {Brady}(2015)}]{takatori_theory_2015}%
  \BibitemOpen
  \bibfield  {author} {\bibinfo {author} {\bibfnamefont {S.}~\bibnamefont
  {Takatori}}\ and\ \bibinfo {author} {\bibfnamefont {J.}~\bibnamefont
  {Brady}},\ }\href@noop {} {\bibfield  {journal} {\bibinfo  {journal} {Soft
  Matter}\ }\textbf {\bibinfo {volume} {11}},\ \bibinfo {pages} {7920}
  (\bibinfo {year} {2015})}\BibitemShut {NoStop}%
\bibitem [{\citenamefont {Pessot}\ \emph {et~al.}(2018)\citenamefont {Pessot},
  \citenamefont {Löwen},\ and\ \citenamefont {Menzel}}]{pessot18}%
  \BibitemOpen
  \bibfield  {author} {\bibinfo {author} {\bibfnamefont {G.}~\bibnamefont
  {Pessot}}, \bibinfo {author} {\bibfnamefont {H.}~\bibnamefont {Löwen}},\
  and\ \bibinfo {author} {\bibfnamefont {A.~M.}\ \bibnamefont {Menzel}},\
  }\href@noop {} {\bibfield  {journal} {\bibinfo  {journal} {Molecular
  Physics}\ }\textbf {\bibinfo {volume} {116}},\ \bibinfo {pages} {3401}
  (\bibinfo {year} {2018})}\BibitemShut {NoStop}%
\bibitem [{\citenamefont {B\'ardfalvy}\ \emph {et~al.}(2020)\citenamefont
  {B\'ardfalvy}, \citenamefont {Anjum}, \citenamefont {Nardini}, \citenamefont
  {Morozov},\ and\ \citenamefont {Stenhammar}}]{dora2020}%
  \BibitemOpen
  \bibfield  {author} {\bibinfo {author} {\bibfnamefont {D.}~\bibnamefont
  {B\'ardfalvy}}, \bibinfo {author} {\bibfnamefont {S.}~\bibnamefont {Anjum}},
  \bibinfo {author} {\bibfnamefont {C.}~\bibnamefont {Nardini}}, \bibinfo
  {author} {\bibfnamefont {A.}~\bibnamefont {Morozov}},\ and\ \bibinfo {author}
  {\bibfnamefont {J.}~\bibnamefont {Stenhammar}},\ }\href@noop {} {\bibfield
  {journal} {\bibinfo  {journal} {Phys. Rev. Lett.}\ }\textbf {\bibinfo
  {volume} {125}},\ \bibinfo {pages} {018003} (\bibinfo {year}
  {2020})}\BibitemShut {NoStop}%
\bibitem [{\citenamefont {Brotto}\ \emph {et~al.}(2015)\citenamefont {Brotto},
  \citenamefont {Bartolo},\ and\ \citenamefont {Saintillan}}]{Brotto15}%
  \BibitemOpen
  \bibfield  {author} {\bibinfo {author} {\bibfnamefont {T.}~\bibnamefont
  {Brotto}}, \bibinfo {author} {\bibfnamefont {D.}~\bibnamefont {Bartolo}},\
  and\ \bibinfo {author} {\bibfnamefont {D.}~\bibnamefont {Saintillan}},\
  }\href@noop {} {\bibfield  {journal} {\bibinfo  {journal} {J. Nonlinear
  Sci.}\ }\textbf {\bibinfo {volume} {25}},\ \bibinfo {pages} {1125–1139}
  (\bibinfo {year} {2015})}\BibitemShut {NoStop}%
\bibitem [{\citenamefont {Batchelor}(1970)}]{batchelor70}%
  \BibitemOpen
  \bibfield  {author} {\bibinfo {author} {\bibfnamefont {G.}~\bibnamefont
  {Batchelor}},\ }\href@noop {} {\bibfield  {journal} {\bibinfo  {journal} {J.
  Fluid Mech.}\ }\textbf {\bibinfo {volume} {44}},\ \bibinfo {pages} {419}
  (\bibinfo {year} {1970})}\BibitemShut {NoStop}%
\bibitem [{\citenamefont {Keller}\ and\ \citenamefont
  {Rubinow}(1976)}]{keller76}%
  \BibitemOpen
  \bibfield  {author} {\bibinfo {author} {\bibfnamefont {J.}~\bibnamefont
  {Keller}}\ and\ \bibinfo {author} {\bibfnamefont {S.}~\bibnamefont
  {Rubinow}},\ }\href@noop {} {\bibfield  {journal} {\bibinfo  {journal} {J.
  Fluid Mech.}\ }\textbf {\bibinfo {volume} {75}},\ \bibinfo {pages} {705}
  (\bibinfo {year} {1976})}\BibitemShut {NoStop}%
\bibitem [{\citenamefont {Simha}\ and\ \citenamefont
  {Ramaswamy}(2002)}]{simha02}%
  \BibitemOpen
  \bibfield  {author} {\bibinfo {author} {\bibfnamefont {R.}~\bibnamefont
  {Simha}}\ and\ \bibinfo {author} {\bibfnamefont {S.}~\bibnamefont
  {Ramaswamy}},\ }\href@noop {} {\bibfield  {journal} {\bibinfo  {journal}
  {Phys. A}\ }\textbf {\bibinfo {volume} {306}},\ \bibinfo {pages} {262}
  (\bibinfo {year} {2002})}\BibitemShut {NoStop}%
\bibitem [{\citenamefont {Jeffery}(1922)}]{jeffery22}%
  \BibitemOpen
  \bibfield  {author} {\bibinfo {author} {\bibfnamefont {G.}~\bibnamefont
  {Jeffery}},\ }\href@noop {} {\bibfield  {journal} {\bibinfo  {journal} {Proc.
  Roy. Soc. Lond. Ser. A}\ }\textbf {\bibinfo {volume} {102}},\ \bibinfo
  {pages} {161} (\bibinfo {year} {1922})}\BibitemShut {NoStop}%
\bibitem [{\citenamefont {Doi}\ and\ \citenamefont {Edwards}(1988)}]{doi88}%
  \BibitemOpen
  \bibfield  {author} {\bibinfo {author} {\bibfnamefont {M.}~\bibnamefont
  {Doi}}\ and\ \bibinfo {author} {\bibfnamefont {S.}~\bibnamefont {Edwards}},\
  }\href@noop {} {\emph {\bibinfo {title} {The theory of polymer dynamics}}}\
  (\bibinfo  {publisher} {Oxford University Press, {USA}},\ \bibinfo {year}
  {1988})\BibitemShut {NoStop}%
\bibitem [{\citenamefont {Saintillan}\ and\ \citenamefont
  {Shelley}(2012)}]{SS2012}%
  \BibitemOpen
  \bibfield  {author} {\bibinfo {author} {\bibfnamefont {D.}~\bibnamefont
  {Saintillan}}\ and\ \bibinfo {author} {\bibfnamefont {M.}~\bibnamefont
  {Shelley}},\ }\href@noop {} {\bibfield  {journal} {\bibinfo  {journal} {J.
  Royal Soc. Interface}\ }\textbf {\bibinfo {volume} {9}},\ \bibinfo {pages}
  {571} (\bibinfo {year} {2012})}\BibitemShut {NoStop}%
\bibitem [{\citenamefont {Yan}\ \emph {et~al.}(2019)\citenamefont {Yan},
  \citenamefont {Zhang},\ and\ \citenamefont {Shelley}}]{yan2019}%
  \BibitemOpen
  \bibfield  {author} {\bibinfo {author} {\bibfnamefont {W.}~\bibnamefont
  {Yan}}, \bibinfo {author} {\bibfnamefont {H.}~\bibnamefont {Zhang}},\ and\
  \bibinfo {author} {\bibfnamefont {M.}~\bibnamefont {Shelley}},\ }\href@noop
  {} {\bibfield  {journal} {\bibinfo  {journal} {J. Chem. Phys.}\ }\textbf
  {\bibinfo {volume} {150}},\ \bibinfo {pages} {064109} (\bibinfo {year}
  {2019})}\BibitemShut {NoStop}%
\bibitem [{\citenamefont {Wajnryb}\ \emph {et~al.}(2013)\citenamefont
  {Wajnryb}, \citenamefont {Mizerski}, \citenamefont {Zuk},\ and\ \citenamefont
  {Szymczak}}]{wajnryb13}%
  \BibitemOpen
  \bibfield  {author} {\bibinfo {author} {\bibfnamefont {E.}~\bibnamefont
  {Wajnryb}}, \bibinfo {author} {\bibfnamefont {K.}~\bibnamefont {Mizerski}},
  \bibinfo {author} {\bibfnamefont {P.}~\bibnamefont {Zuk}},\ and\ \bibinfo
  {author} {\bibfnamefont {P.}~\bibnamefont {Szymczak}},\ }\href@noop {}
  {\bibfield  {journal} {\bibinfo  {journal} {J. Fluid Mech.}\ }\textbf
  {\bibinfo {volume} {731}},\ \bibinfo {pages} {R3} (\bibinfo {year}
  {2013})}\BibitemShut {NoStop}%
\bibitem [{\citenamefont {Ying}\ \emph {et~al.}(2004)\citenamefont {Ying},
  \citenamefont {Biros},\ and\ \citenamefont {Zorin}}]{ying04}%
  \BibitemOpen
  \bibfield  {author} {\bibinfo {author} {\bibfnamefont {L.}~\bibnamefont
  {Ying}}, \bibinfo {author} {\bibfnamefont {G.}~\bibnamefont {Biros}},\ and\
  \bibinfo {author} {\bibfnamefont {D.}~\bibnamefont {Zorin}},\ }\href@noop {}
  {\bibfield  {journal} {\bibinfo  {journal} {J. Comput. Phys}\ }\textbf
  {\bibinfo {volume} {196}},\ \bibinfo {pages} {591} (\bibinfo {year}
  {2004})}\BibitemShut {NoStop}%
\bibitem [{\citenamefont {Yan}\ and\ \citenamefont
  {Shelley}(2018)}]{yan2018214}%
  \BibitemOpen
  \bibfield  {author} {\bibinfo {author} {\bibfnamefont {W.}~\bibnamefont
  {Yan}}\ and\ \bibinfo {author} {\bibfnamefont {M.}~\bibnamefont {Shelley}},\
  }\href@noop {} {\bibfield  {journal} {\bibinfo  {journal} {J. Comput. Phys.}\
  }\textbf {\bibinfo {volume} {355}},\ \bibinfo {pages} {214} (\bibinfo {year}
  {2018})}\BibitemShut {NoStop}%
\bibitem [{\citenamefont {Saad}\ and\ \citenamefont {Schultz}(1986)}]{Saad86}%
  \BibitemOpen
  \bibfield  {author} {\bibinfo {author} {\bibfnamefont {Y.}~\bibnamefont
  {Saad}}\ and\ \bibinfo {author} {\bibfnamefont {Y.}~\bibnamefont {Schultz}},\
  }\href@noop {} {\bibfield  {journal} {\bibinfo  {journal} {SIAM J. Sci. Stat.
  Comput.}\ }\textbf {\bibinfo {volume} {7}},\ \bibinfo {pages} {856} (\bibinfo
  {year} {1986})}\BibitemShut {NoStop}%
\bibitem [{\citenamefont {Barnett}\ \emph {et~al.}(2019)\citenamefont
  {Barnett}, \citenamefont {Magland},\ and\ \citenamefont
  {af~Klinteberg}}]{barnett19}%
  \BibitemOpen
  \bibfield  {author} {\bibinfo {author} {\bibfnamefont {A.~H.}\ \bibnamefont
  {Barnett}}, \bibinfo {author} {\bibfnamefont {J.}~\bibnamefont {Magland}},\
  and\ \bibinfo {author} {\bibfnamefont {L.}~\bibnamefont {af~Klinteberg}},\
  }\href@noop {} {\bibfield  {journal} {\bibinfo  {journal} {SIAM Journal on
  Scientific Computing}\ }\textbf {\bibinfo {volume} {41}},\ \bibinfo {pages}
  {C479} (\bibinfo {year} {2019})}\BibitemShut {NoStop}%
\bibitem [{\citenamefont {{\ifmmode \check{S}\else \v{S}}\fi{}kult\'ety}\ \emph
  {et~al.}(2020)\citenamefont {{\ifmmode \check{S}\else \v{S}}\fi{}kult\'ety},
  \citenamefont {Nardini}, \citenamefont {Stenhammar}, \citenamefont
  {Marenduzzo},\ and\ \citenamefont {Morozov}}]{Viktor20}%
  \BibitemOpen
  \bibfield  {author} {\bibinfo {author} {\bibfnamefont {V.}~\bibnamefont
  {{\ifmmode \check{S}\else \v{S}}\fi{}kult\'ety}}, \bibinfo {author}
  {\bibfnamefont {C.}~\bibnamefont {Nardini}}, \bibinfo {author} {\bibfnamefont
  {J.}~\bibnamefont {Stenhammar}}, \bibinfo {author} {\bibfnamefont
  {D.}~\bibnamefont {Marenduzzo}},\ and\ \bibinfo {author} {\bibfnamefont
  {A.}~\bibnamefont {Morozov}},\ }\href@noop {} {\bibfield  {journal} {\bibinfo
   {journal} {Phys. Rev. X}\ }\textbf {\bibinfo {volume} {10}},\ \bibinfo
  {pages} {031059} (\bibinfo {year} {2020})}\BibitemShut {NoStop}%
\end{thebibliography}%
\bibliographystyle{apsrev4-2} 

\end{document}